\documentclass[conference]{IEEEtran}

\IEEEoverridecommandlockouts
\usepackage{cite}
\usepackage{amsmath,amssymb,amsfonts,balance}
\usepackage{algorithmic}
\usepackage{graphicx}
\usepackage{textcomp}
\usepackage{xcolor}
\usepackage{ucs}
\usepackage[utf8x]{inputenc}
\usepackage{cite, amsfonts, amssymb, amsthm, bm, bbm, graphicx, relsize, multirow, booktabs, tikz,subfigure,soul}
\usepackage{stackrel}
\usepackage[american]{babel}
\usepackage[T1]{fontenc}
\usepackage{algorithmic, algorithm}
\usepackage[multiple]{footmisc}

\newcommand{\ds}{\displaystyle}
\newcommand{\norm}[1]{\left\lVert#1\right\rVert}
\def\BibTeX{{\rm B\kern-.05em{\sc i\kern-.025em b}\kern-.08em
    T\kern-.1667em\lower.7ex\hbox{E}\kern-.125emX}}
\begin{document}
\bstctlcite{IEEEexample:BSTcontrol}
\title{Transmit Power Allocation for Joint Communication and Sensing through Massive MIMO Arrays
}

\author{\IEEEauthorblockN{Stefano Buzzi}
\IEEEauthorblockA{\textit{DIEI} \\
\textit{University of Cassino} \\ \textit{and Southern Latium}\\
Cassino, Italy \\
{ \tt buzzi@unicas.it}}
\and
\IEEEauthorblockN{Carmen D'Andrea}
\IEEEauthorblockA{\textit{DIEI} \\
\textit{University of Cassino} \\ \textit{and Southern Latium}\\
Cassino, Italy \\
{\tt carmen.dandrea@unicas.it}}
\and
\IEEEauthorblockN{Marco Lops}
\IEEEauthorblockA{\textit{DIETI} \\
\textit{University ``Federico II'' of Naples}\\
Naples, Italy \\
{\tt lops@unina.it }}
}

\maketitle

\begin{abstract}
The paper considers a scenario where a  base station (BS), equipped with a large-scale antenna array,  execute, using the same frequency range, both communication with mobile users and radar surveillance of the surrounding environment, relying on the ability of the massive MIMO array to synthesize multiple narrow beams. Based on an OFDM signaling format for both communication and surveillance tasks, a lower bound to the system achievable downlink rate is provided, along with a GLRT detection rule that does not require any knowledge about the target parameters. Then, a power allocation strategy is proposed, aimed at maximizing the fairness across the mobile users, while guaranteeing a minimum signal to interference ratio threshold value for the radar system. Numerical results show that the system performs effectively, and that the power control procedure helps in improving the system fairness.
\end{abstract}


\section{Introduction}
Recently, there is increasing attention on the topic of the co-existence, in the same frequency band, of both radar and communication systems: see \cite{ZhengSPM} for a recent review of the progress in this area. 
The interest in this field is justified by the progressive scaling up of frequency bands, traditionally used in  radar systems, produced by the standard evolution of the cellular networks from GSM to the fifth generation (5G). Most of the work in this area has focused on the case in which the radar system and the communication system are distinct \cite{JSTSP, DAndrea_TWC2019}, and has considered several degrees of cooperation, ranging from totally uncoordinated design of the two systems to the case of full cooperation. In the recent paper \cite{Buzzi_Asilomar2019}, instead, the authors have considered the case in which a base station (BS), that we nickname as radar-BS, relying on a shared large antenna array, performs both the communication and radar sensing tasks, using co-located transceivers for both functions. The paper was inspired by  \cite{Caire_OTFS_ICC2019} where a similar scenario was considered with reference to a vehicular radar and communication system. The working assumption of 
\cite{Buzzi_Asilomar2019} is that the  massive antenna array can both operate as a MIMO radar with co-located antennas -- transmitting radar signals pointing at positive elevation angles -- and perform signal-space beamforming to communicate with users mainly based on the ground. It is anticipated that a radar-BS may turn out to be extremely popular and useful in the near future, when we expect that several unmanned flying objects will populate the sky above our heads, and it will thus be critical to be able to safely control and track them. 

While in  \cite{Buzzi_Asilomar2019}  the benefit of power control strategies has not been investigating, this is the main goal of this paper. Indeed,  a power allocation strategy aimed at maximizing the system fairness across users of the communication system, subject to a minimum signal to interference ratio (SIR) constraint for the radar system is developed here. Our results will show that the power allocation strategy offers good performance in term of minimum rate for the users of the communication system while maintaining also good performance in term of the detection probability of the radar. 

The paper is organized as follows. Next section contains the description of the considered scenario and of the channel and signal models. Section \ref{Downlink_performance} is devoted to the description of the  downlink achievable rate lower bound, while in Section \ref{Power_all_section} the power allocation strategy is derived. Numerical results are discussed  in Section \ref{Numerica_results}, while, finally, concluding remarks are given in Section \ref{Conclusions}. 
\begin{figure}
\begin{center}
\includegraphics[scale=1.15]{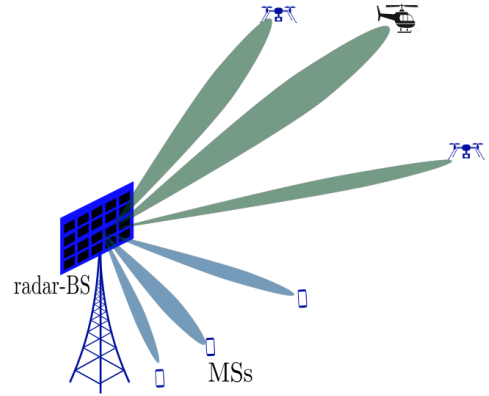}
\end{center}
\caption{Representation of the considered scenario.}
\label{Fig:scenario}
\end{figure}
\section{System model} \label{System_model}
We consider the scenario  depicted in Fig. \ref{Fig:scenario}. A radar-BS equipped with a large-scale planar antenna array with $N_A=N_{A,y}N_{A,z}$ elements  ($N_{A,y}$ on the horizontal axis and $N_{A,z}$ on the vertical axis), jointly serves $K$ single-antenna mobile stations and performs surveillance tasks of the surrounding space -- through electronically steered phased-array beams pointed at  positive elevation angles -- using the same frequency range. The time-division-duplex (TDD) protocol is used for data communication with the mobile stations, so as to exploit the uplink/downlink channel reciprocity.
We denote by $B$ the total bandwidth and by $f_c$ the carrier frequency. Orthogonal 
frequency division multiplexing (OFDM) modulation is used for both communication and surveillance tasks; the total bandwidth is thus divided into $M$ subcarriers, i.e. $B=M \Delta f$, where $\Delta f$ denotes the subcarrier bandwidth.

\subsection{Channel model} \label{Channel_model_section}
We first provide the model for the channel between the BS and the potential target. 
Assume that a target with radial speed  $v$ [m/sec] with respect to the radar-BS is present in the surveillance area. 
The channel from the BS to the target and then, upon reflection, again to the BS is modeled as a random linear time variant system with matrix-valued channel impulse response expressed as 
\begin{equation}
\widetilde{\mathbf{H}}_T(t,\tau)=\mathbf{H}_T \delta(t-\tau)e^{j2\pi\nu t} \; .
\label{radar_channel}
\end{equation}
In \eqref{radar_channel}, 
$\tau$ and $\nu$ denote the round-trip delay and the Doppler shift induced by the target speed; moreover, letting the pair $(\phi, \theta)$ denote the azimuth and elevation angles of the target with respect to the BS antenna, we have 
$
\mathbf{H}_T= \alpha_T \mathbf{a}\left(\phi,\theta\right) \mathbf{a}\left(\phi,\theta\right)^H,
$
with $\alpha_T$ a complex coefficient taking into account the  target reflection coefficient and the path-loss. The vector  $\mathbf{a}\left(\phi,\theta\right)$ represents the BS antenna array response vector associated with the angles $(\phi, \theta)$, i.e.,
\begin{equation}
\begin{array}{lllll}
\mathbf{a}\left(\phi,\theta\right)&=\left[1, \ldots, e^{-j\tilde{k}d\left(a_y \sin(\phi)\sin(\theta) + a_z \cos(\theta)\right)}, \right. \\ & \left. \ldots, e^{-j\tilde{k}d\left((N_{A,y}-1) \sin(\phi)\sin(\theta) + (N_{A,z}-1) \cos(\theta) \right)} \right] 
\end{array}
\label{array_response}
\end{equation}
with $\tilde{k}=2\pi/\lambda$  the wavenumber, $\lambda$ the wavelength and $d$ the inter-element spacing.

With regard to the channel between the radar-BS and the generic $k$-th user,
$\mathbf{h}_k$ say, three different scenarios will be considered: Rayleigh-distributed channel, pure line-of-sight (LoS) channel with uniform phase, and Rice-distributed channels. For the Rayleigh case, we have
\begin{equation}
\mathbf{h}_k= \sqrt{\beta_k} \mathbf{g}_k \, ,
\label{Rayleigh_channel}
\end{equation}
where $\beta_k$ subsumes the path-loss and the shadow fading coefficient, and $\mathbf{g}_k \sim \mathcal{CN}\left(\mathbf{0}, \mathbf{I}_{N_A}\right)$. 
If the LoS channel model is in force, we have 
\begin{equation}
\mathbf{h}_k= \sqrt{\beta_k} e^{j \psi_k} \mathbf{a}\left(\varphi_k,\vartheta_k\right) \, ,
\label{LOS_channel}
\end{equation}
with $\beta_k$ representing the path-loss, $\psi_k$ is  the random phase uniformly distributed in $[0, 2\pi]$ and $\mathbf{a}\left(\varphi_k,\vartheta_k\right)$ is the BS antenna array response  evaluated at the azimuth and elevation angles, $\left(\varphi_k,\vartheta_k\right)$ say, of the $k$-th user.
Finally, for the  Rice-distributed  channel we have
\begin{equation}
\mathbf{h}_k= \sqrt{\frac{\beta_k}{K_{k}+1}} \left[ \sqrt{K_k} e^{j \psi_k} \mathbf{a}\left(\varphi_k,\vartheta_k\right) +\mathbf{g}_k \right]\, ,
\label{Rice_channel}
\end{equation}
where the Ricean $K$-factor is
\begin{equation}
K_k=\frac{p_{\rm LoS}(d_{k, {\rm 2D}})}{1-p_{\rm LoS}(d_{k, {\rm 2D}})} \, ,
\label{K_factors}
\end{equation}
$d_{k, {\rm 2D}}$ is the 2D distance between the BS and the  $k$-th user, and $p_{\rm LoS}(d_{k, {\rm 2D}})$ is the LoS probability.

\subsection{Signal model}

Following \cite{Caire_OTFS_ICC2019,Buzzi_Asilomar2019}, we assume that a standard cyclic prefix (CP) OFDM modulation is used for both the communication and radar surveillance tasks, with $\Delta f$ the subcarrier spacing. 
Let $T_0=T_{ \rm CP} +T_{\rm s}$ be the OFDM symbol duration, with $T_{ \rm CP}$ and $T_{\rm s}=1/\Delta f$ denoting the CP and the symbol duration, respectively. The OFDM frame duration is $T_{\rm OFDM}=N T_0$. The unit-power data symbols intended for the $k$-th user are denoted by $x_k(n,m)$ for $n=0,\ldots, N-1$,  $m=0,\ldots, M-1$, and  are arranged in a $N \times M$ grid. 
Similarly, the fictitious unit-power symbols used for radar detection are denoted by  $x_R(n,m)$ and arranged in a $N \times M$ grid. The continuous-time OFDM signals with CP intended to the $k$-th user  and intended for radar surveillance can be thus written as
\begin{equation}
s_k(t)=\ds \sum_{n=0}^{N-1} \sum_{m=0}^{M-1} x_k(n,m) \text{rect} (t-n T_0) e^{j2\pi m \Delta f (t- T_{\rm CP}-nT_0)},
\label{signal_k}
\end{equation}
and
\begin{equation}
s_R(t)=\ds \sum_{n=0}^{N-1} \sum_{m=0}^{M-1} x_R(n,m) \text{rect} (t-n T_0) e^{j2\pi m \Delta f (t- T_{\rm CP}-nT_0)},
\label{signal_R}
\end{equation}
respectively, with  $\text{rect}(t)$ a rectangular pulse supported on $[0, T_0]$. 
Accordingly,
denoting by $\eta_k$ the power used by the radar-BS to transmit to the $k$-th 
user and $\eta_R$ the power used for surveillance purposes on each symbol of the $N \times M$ grid, 
the $N_A$-dimensional signal transmitted by the radar-BS can be shown to be written as
\begin{equation}
\mathbf{s}(t)=\ds \sum_{k=1}^K { \sqrt{\eta_k} s_k(t) \mathbf{w}_k } + \sqrt{\eta_R} s_R(t) \mathbf{w}_R\left(\phi,\theta\right) \, .
\label{signal_DL_st}
\end{equation}
In \eqref{signal_DL_st},  $\mathbf{w}_k$ is the beamforming vector used to transmit to the $k$-th user, while $\mathbf{w}_R\left(\phi,\theta\right)$ is the beamforming vector for surveillance tasks in the  direction corresponding to the azimuth and elevation angles $(\phi,\theta)$. 
Two possible choices are considered in this paper for the radar beamforming vector  $\mathbf{w}_R\left(\phi,\theta\right)$. 
The former is to use the radar-BS antenna as a phased array producing a phased beam towards the direction $(\phi,\theta)$, i.e.:
\begin{equation}
\mathbf{w}_R\left(\phi,\theta\right)= \frac{1}{\sqrt{N_A}}\mathbf{a}\left(\phi,\theta\right)\, .
\label{pure_radar_direction}
\end{equation}
The above choice would however cause some interference to ground users; an alternative is thus to modify the beamformer in \eqref{pure_radar_direction} in order to force to zero the interference produced by the radar signal to the mobile users. 
Letting $\widetilde{\mathbf{U}}$ denote a matrix whose columns form a basis for the subspace spanned by the estimated channel vectors $\left[  \widehat{\mathbf{h}}_1, \ldots,  \widehat{\mathbf{h}}_K \right]$, we have thus the zero-forcing radar (ZFR) beamformer:
\begin{equation}
\mathbf{w}_R\left(\phi,\theta\right)= \frac{\left( \mathbf{I}_{N_A}- \widetilde{\mathbf{U}} \widetilde{\mathbf{U}}^H\right) \mathbf{a}\left(\phi,\theta\right)}{\norm{\left( \mathbf{I}_{N_A}- \widetilde{\mathbf{U}} \widetilde{\mathbf{U}}^H \right) \mathbf{a}\left(\phi,\theta\right)}}\, .
\label{ZF_radar}
\end{equation}
Two comments are in order about the beamformer \eqref{ZF_radar}. First of all, the above equation implicitly assumes that $N_A>K$, i.e. the number of antennas at the radar-BS much be larger than the number of users in order to be able to null the beamformer projection along $K$ signal space directions. Second, the ZFR beamformer is able to actually null to zero the interference from the radar signal to the mobile users only under the assumption of perfect channel state information; in practice, only a fraction of this interference will be canceled, depending on the accuracy of the channel estimates.

\begin{figure*}
\begin{small}
\begin{equation}
\begin{array}{ll}
\mathcal{R}_k^{\rm PM}= B\ds \frac{\tau_d}{\tau_c}\log_2 \left( 1+
\ds \frac{\eta_k \gamma_k}
{\ds \sum_{j=1}^K {\frac{\eta_j}{\eta_{{\rm p},j}} \left( \frac{\text{tr}\left( \mathbf{R}_{y,j} \overline{\mathbf{H}}_k \right) }{\gamma_j} + \eta_{{\rm p},k} \frac{\delta_k}{\gamma_j} \left|\boldsymbol{\phi}_k^H \boldsymbol{\phi}_j \right|^2\right) } - \eta_k \gamma_k +\eta_R \mathbf{w}_R^H\left(\phi,\theta\right) \overline{\mathbf{H}}_k \mathbf{w}_R\left(\phi,\theta\right) + \sigma^2_z }\right) 
\end{array}
\label{SE_DL_PM}
\end{equation}
\begin{equation}
\begin{array}{ll}
\mathcal{R}_k^{\rm LMMSE}= B\ds \frac{\tau_d}{\tau_c}\log_2 \left( 1+
\ds \frac{\eta_k \widetilde{\gamma}_k}
{ \ds \sum_{j=1}^K {\eta_j \left(\sqrt{\eta_{{\rm p},j}} \frac{\text{tr}\left(  \overline{\mathbf{H}}_j\mathbf{E}_{j} \overline{\mathbf{H}}_k \right)}{\widetilde{\gamma}_j}  + \eta_{{\rm p},k} \frac{\widetilde{\delta}_j^{(k)}}{\widetilde{\gamma}_j} \left|\boldsymbol{\phi}_k^H \boldsymbol{\phi}_j \right|^2 \right) } - \eta_k \widetilde{\gamma}_k +\eta_R \mathbf{w}_R^H\left(\phi,\theta\right) \overline{\mathbf{H}}_k \mathbf{w}_R\left(\phi,\theta\right)+ \sigma^2_z }\right) 
\end{array}
\label{SE_DL_MMSE}
\end{equation}
\hrulefill
\end{small}
\end{figure*}

\section{Transceiver processing and downlink performance analysis} \label{Downlink_performance}

We now detail the transceiver processing for the channel estimation and downlink data transmission phases.

\subsection{Uplink channel estimation}
Since the BS does not transmit during this phase, the received signal will not contain any possible target echo.
Let us denote by $\tau_c$ the dimension in time/frequency samples of the channel coherence length, and by $\tau_p < \tau_c$ the dimension of the uplink training phase. 
We also denote by $\boldsymbol{\phi}_k \in \mathbb{C}^{\tau_p}$ the pilot sequence transmitted by the $k$-th user, with $\|\boldsymbol{\phi}_k\|^2=1\, , \forall k$. 
Based on the above assumptions, the signal received at the radar-BS during the training phase 
 can be therefore expressed as the following $({N_{A}  \times \tau_p})$-dimensional matrix:
\begin{equation}
\mathbf{Y}_{\rm p} = \ds \sum_{k=1}^K \ds \sqrt{\eta_{{\rm p},k}} \mathbf{h}_{k}\boldsymbol{\phi}_k^H + \mathbf{W}_{\rm p} \; ,
\label{eq:y_pilot}
\end{equation}
with $\eta_{{\rm p},k}$ denoting the $k$-th user transmitted power, and $\mathbf{W}_{\rm p} \in \mathbb{C}^{N_A  \times \tau_p}$ represents the thermal noise contribution and out-of-cell interference at the radar-BS. The entries of  $\mathbf{W}_{\rm p}$  are modeled as i.i.d.  ${\cal CN}(0, \sigma^2_w)$ RVs. Given the 
observable $\mathbf{Y}_{\rm p}$ reported in \eqref{eq:y_pilot}, the radar-BS forms the statistics 
$\mathbf{y}_{{\rm p},k}=\mathbf{Y}_{\rm p} \boldsymbol{\phi}_k$, $\forall \; k=1,\ldots, K$.
In order to estimate the channel vectors 
$\mathbf{h}_{k}, \forall \; k=1,\ldots, K$, two possible channel estimation (CE) techniques will be considered: pilot matched CE (PM-CE) and  linear minimum-mean-square-error CE (LMMSE-CE). 
 
 For the case of PM-CE, the channel estimate of $\mathbf{h}_k$ is written as
\begin{equation}
 \widehat{\mathbf{h}}_k=\frac{1}{\sqrt{\eta_{{\rm p},k}}}\mathbf{y}_{{\rm p},k} \, .
\end{equation}

For LMMSE-CE, instead, the channel estimate can be shown to be written as \cite{kay1993fundamentals} 
\begin{equation}
 \widehat{\mathbf{h}}_k=\mathbf{E}_k^H\mathbf{y}_{{\rm p},k} \, ,
\end{equation}
where
\begin{equation*}
\begin{array}{llll}
&\mathbf{E}_k= \sqrt{\eta_{{\rm p},k}} \mathbf{R}_{y,k}^{-1} \overline{\mathbf{H}}_k \, , \\
&\mathbf{R}_{y,k}= \sum_{i=1}^K \ds \eta_{{\rm p},i} \overline{\mathbf{H}}_i\left|\boldsymbol{\phi}_i^H \boldsymbol{\phi}_k \right|^2+ \sigma^2_w \mathbf{I}_{N_A} \, ,
\end{array}
\end{equation*}
and $\overline{\mathbf{H}}_k$ is an $(N_A \times N_A)$-dimensional matrix depending on the adopted channel model. For the case of Rayleigh-distributed channel, Eq. \eqref{Rayleigh_channel}, we have
$
\overline{\mathbf{H}}_k= \beta_k \mathbf{I}_{N_A} \, ;
$
for the case of LoS channel, Eq. \eqref{LOS_channel}, we have
$
\overline{\mathbf{H}}_k= \beta_k \mathbf{a}\left(\varphi_k,\vartheta_k\right) \mathbf{a}^H\left(\varphi_k,\vartheta_k\right) \, ,
$
while finally, for the case of Rice-distributed channel, Eq. \eqref{Rice_channel}, we have
\begin{equation}
\overline{\mathbf{H}}_k= \frac{\beta_k}{K_{k}+1}\left[ K_{k} \mathbf{a}\left(\varphi_k,\vartheta_k\right) \mathbf{a}^H\left(\varphi_k,\vartheta_k\right) + \mathbf{I}_{N_A} \right]\, .
\label{H_bar_matrix_Rice}
\end{equation}

\subsection{Downlink data transmission}
On the downlink, the signal received by the $k$-th user  is expressed in discrete-time as follows:
\begin{equation}
\begin{array}{llll}
y_k(n,m)=&\sqrt{\eta_k} \mathbf{h}_k^H \mathbf{w}_k x_k(n,m) + \ds \sum_{\substack{j=1 \\ j \neq k}}^K {\sqrt{\eta_j} \mathbf{h}_k^H \mathbf{w}_j x_j(n,m)}  \\& + \sqrt{\eta_R} \mathbf{h}_k^H \mathbf{w}_R\left(\phi,\theta\right) x_R(n,m) + z_k(n,m) \, ,
\end{array}
\label{DL_signal_user}
\end{equation}
where $z_k(n,m)\sim {\cal CN}(0, \sigma^2_z)$ is the AWGN contribution. The quantity $y_k(n,m)$ thus represents the soft estimate of the information symbol $x_k(n,m)$ and can be further processed for data detection. 

Regarding the system performance analysis,  starting from Eq. \eqref{DL_signal_user}, and exploiting the use-and-then-forget bounding technique \cite{marzetta2016fundamentals},  the closed form achievable rate formulas, reported 
in Eqs. \eqref{SE_DL_PM} and \eqref{SE_DL_MMSE} at the top of next page, can be derived 
for the PM-CE and for the LMMSE-CE, assuming channel matched beamforming, i.e., $\mathbf{w}_k=\widehat{\mathbf{h}}_k/\norm{\widehat{\mathbf{h}}_k}$, respectively. In these expressions,  $\tau_d=\tau_c-\tau_p$ is the dimension in time/frequency samples of the downlink data transmission phase, 
$
\gamma_k=\text{tr}\left(\overline{\mathbf{H}}_k\right)$, and $\widetilde{\gamma}_k=\sqrt{\eta_{{\rm p},k}}\text{tr}\left(\overline{\mathbf{H}}_k\mathbf{E}_k\right).
$ Moreover, for the case of Rayleigh channel, we have
$
\delta_k= \beta_k^2 N_A^2 \; \; \text{and} \; \; \widetilde{\delta}_j^{(k)}=\beta_k^2\text{tr}\left(\mathbf{E}_j^H\right) 
$;  for the case of LoS channel, we have
$
\delta_k= 0 \; \; \text{and} \; \; \widetilde{\delta}_j^{(k)}=0
\label{delta_LOS}
$;
and, finally, for the case of Rice channel, we have
\begin{equation}
\delta_k= \left( \frac{\beta_k}{K_{k}+1} \right)^2 N_A \left( N_A+ 2 K_{k}\right) \, ,
\label{delta_k_Rice}
\end{equation}

\begin{equation}
\begin{array}{lllll}
\widetilde{\delta}_j^{(k)}&=  \left(  \frac{\beta_k}{K_{k}+1} \right)^2 \left[\text{tr}\left( \mathbf{E}_j^H \right)  \right. \\ & \left. + \ds  2 K_{k} \mathbb{R}\left\lbrace  \text{tr}\left( \mathbf{a}^H\left(\varphi_k,\vartheta_k\right)\mathbf{E}_j^H \mathbf{a}\left(\varphi_k,\vartheta_k\right)  \mathbf{E}_j\right)\right\rbrace  \right].
\end{array}
\end{equation}

\subsection{Radar processing}

The full derivation of the signal processing tasks for the radar is omitted for the sake of brevity. 
In order to perform joint radar detection in the direction defined by the angles $(\phi, \theta)$, given the total ignorance on the potential target reflectivity, distance and doppler frequency, 
upon defining the uniformly-spaced grid in the delay and Doppler domain $\mathcal{G}$, a Generalized Likelihood Ratio Test (GLRT) can be implemented as follows 
\begin{equation}
\max_{\tau, \nu \in \cal{G}} \left|  \ds \sum_{n=0}^{N-1} \sum_{m=0}^{M-1} e^{-j2\pi\nu n T_0} e^{j2\pi m \Delta f \tau} \mathbf{u}(n,m)^H \mathbf{y}(n,m) \right|^2\stackrel[H_0]{H_1}{\gtrless} \gamma
\label{GLRT}
\end{equation}
In the above test, $\mathbf{y}(n,m)$ and  $\mathbf{u}(n,m)$ are $N_A$-dimensional vectors representative of the received and transmitted signals at the radar-BS, respectively.
The reader is referred to \cite{Buzzi_Asilomar2019} for full details about the radar signal processing tasks.

\section{Power allocation} \label{Power_all_section}
The achievable rate lower bound for the $k$-th user in Eqs. \eqref{SE_DL_PM} and \eqref{SE_DL_MMSE} at the top of next page can be compactly written as 
\begin{equation}
\begin{array}{ll}
{\cal R}_k\!= B \ds \frac{\tau_d}{\tau_c}\log_2 \!\!\left( \!\!1+\!\!
\ds \frac{\eta_k \gamma_k}
{\ds \!\!\sum_{j=1}^K {\eta_j \xi_{kj}} + \eta_R \zeta_{k,R}\left(\phi,\theta\right)\! + \!\sigma^2_z }\right) \, ,
\end{array}
\label{Rate_DL_generic}
\end{equation}
where $\tau_d=\tau_c-\tau_p$ is the dimension in time/frequency samples of the downlink data transmission phase. 
The quantities in \eqref{Rate_DL_generic}, for the case of PM channel estimation, can be shown to be expressed as:
\begin{equation}
\begin{array}{llll}
\gamma_k=\text{tr}\left(\overline{\mathbf{H}}_k\right) \, ,
\\
\xi_{kj} = \left\lbrace
\begin{array}{lllll}
\!\!\!\!\ds \frac{1}{\eta_{{\rm p},j}} \left( \frac{\text{tr}\left( \mathbf{R}_{y,j} \overline{\mathbf{H}}_k \right) }{\gamma_j} + \eta_{{\rm p},k} \frac{\delta_k}{\gamma_j} \left|\boldsymbol{\phi}_k^H \boldsymbol{\phi}_j \right|^2\right) \; \text{if} \; j\neq k \\
\!\!\!\!\ds \frac{1}{\eta_{{\rm p},k}} \left( \frac{\text{tr}\left( \mathbf{R}_{y,k} \overline{\mathbf{H}}_k \right) }{\gamma_k} + \eta_{{\rm p},k} \frac{\delta_k}{\gamma_k}\right) \!-\! \gamma_k  \; \text{if} \; j=k
\end{array} \right.
\end{array} \, ,
\end{equation}
while instead, for  the case of MMSE channel estimation, they are written as:
\begin{equation}
\begin{array}{llll}
\gamma_k=\sqrt{\eta_{{\rm p},k}}\text{tr}\left(\overline{\mathbf{H}}_k\mathbf{E}_k\right) \, ,
\\
\xi_{kj} = \left\lbrace
\begin{array}{lllll}
\!\!\!\!\ds \sqrt{\eta_{{\rm p},j}} \frac{\text{tr}\left(  \overline{\mathbf{H}}_j\mathbf{E}_{j} \overline{\mathbf{H}}_k \right)}{\gamma_j}  + \eta_{{\rm p},k} \frac{\widetilde{\delta}_j^{(k)}}{\gamma_j} \left|\boldsymbol{\phi}_k^H \boldsymbol{\phi}_j \right|^2  \; \text{if} \; j\neq k \\
\!\!\!\!\ds \sqrt{\eta_{{\rm p},k}} \frac{\text{tr}\left(  \overline{\mathbf{H}}_k\mathbf{E}_{k} \overline{\mathbf{H}}_k \right)}{\gamma_k}  + \eta_{{\rm p},k} \frac{\widetilde{\delta}_k^{(k)}}{\gamma_k}  \!-\! \gamma_k  \; \text{if} \; j=k
\end{array} \right.
\end{array} \, .
\end{equation}
In the above equations, the quantities $\widetilde{\delta}_j^{(k)}$, $\delta_j$, $\mathbf{R}_{y,j}$, $\mathbf{E}_{j}$ and $\overline{\mathbf{H}}_j$, depend on the adopted channel model and are reported in Section \ref{Downlink_performance}; finally, we have
$
\zeta_{k,R}\left(\phi,\theta\right)= \mathbf{w}_R^H\left(\phi,\theta\right) \overline{\mathbf{H}}_k \mathbf{w}_R\left(\phi,\theta\right)$.

Given the expression of the lower bound achievable rate in Eq. \eqref{Rate_DL_generic}, we formulate the following optimization problem to perform the power allocation:
\begin{subequations}\label{Prob:MinRate}
\begin{align}
&\ds\max_{\boldsymbol{\eta}}\;\min_{1,\ldots, K}\; {\cal R}_k  \left( \boldsymbol{\eta}\right)\label{Prob:aMinRate}\\
&\;\textrm{s.t.}\; \sum_{k=1}^K \eta_{k} + \eta_R \leq \ds \frac{P_{\rm max}}{MN}\;\label{Prob:bMinRate}\\
&\;\;\; \;\;\; \ds \frac{\eta_R \norm{\mathbf{a}\left(\phi,\theta\right)\mathbf{a}^H\left(\phi,\theta\right)\mathbf{w}_R\left(\phi,\theta\right)}^2}{\ds \sum_{k=1}^K \eta_k \norm{\mathbf{a}\left(\phi,\theta\right)\mathbf{a}^H\left(\phi,\theta\right)\mathbf{w}_k}^2 } \geq \rho^*  \label{Prob:cMinRate}
\end{align}
\end{subequations}
where $\boldsymbol{\eta}=\left[ \eta_R,\eta_1,\ldots,\eta_K \right]^T$, $P_{\rm max}$ is the maximum power transmitted from the radar-BS and $\rho^*$ is the signal-to-interference-ratio (SIR) constraint for the radar task. 

Given the monotonicity of $\log_2(\cdot)$, the objective function of  \eqref{Prob:MinRate} can be equivalently rewritten as

\begin{equation}\label{Prob:MinRate2}
\frac{\eta_k \gamma_k}
{\ds \!\!\sum_{j=1}^K {\eta_j \xi_{kj}} + \eta_R \zeta_{k,R}\left(\phi,\theta\right)\! + \!\sigma^2_z }
\end{equation}

Expression \eqref{Prob:MinRate2} is quasi-concave, and so the corresponding optimization problem is quasi-concave. Problem \eqref{Prob:MinRate} can be thus equivalently
reformulated as

\begin{subequations}\label{Prob:MinRate3}
 \begin{align}
&\ds\max_{\boldsymbol{\eta},t }\; \; \; \;  t  \label{Prob:a1MinRate3}\\
&\;\textrm{s.t.}\; \;\frac{\eta_k \gamma_k}
{\ds \!\!\sum_{j=1}^K {\eta_j \xi_{kj}} + \eta_R \zeta_{k,R}\left(\phi,\theta\right)\! + \!\sigma^2_z } \geq t \; \forall k \label{Prob:aMinRate3}\\
& \;\;\; \;\;\;  \sum_{k=1}^K \eta_{k} + \eta_R \leq \ds \frac{P_{\rm max}}{MN}\;\label{Prob:bMinRate3}\\
&\;\;\; \;\;\; \ds \frac{\eta_R \norm{\mathbf{a}\left(\phi,\theta\right)\mathbf{a}^H\left(\phi,\theta\right)\mathbf{w}_R\left(\phi,\theta\right)}^2}{\ds \sum_{k=1}^K \eta_k \norm{\mathbf{a}\left(\phi,\theta\right)\mathbf{a}^H\left(\phi,\theta\right)\mathbf{w}_k}^2 } \geq \rho^*  \label{Prob:cMinRate3}
\end{align}
\end{subequations}

Problem \eqref{Prob:MinRate3} can be solved efficiently by a bisection
search, in each step solving a sequence of convex feasibility problems \cite{boyd2004convex} as detailed in Algorithm \ref{Alg_PowerControl}.

\begin{algorithm}[!t]

\caption{Bisection Algorithm for Solving Problem \eqref{Prob:MinRate3}}

\begin{algorithmic}[1]

\label{Alg_PowerControl}
\STATE  Choose the initial values of $t_{\rm min}$ and $t_{\rm max}$ defining a range of relevant values of the objective function in \eqref{Prob:MinRate3}. Choose a tolerance $\epsilon > 0$.

\WHILE {$t_{\rm max}-t_{\rm min}<\epsilon$}

\STATE Set $ t= \frac{t_{\rm max}+t_{\rm min}}{2} $

\STATE Solve the following convex feasibility program:

\begin{equation}
\left\lbrace
\begin{array}{llll}
&\frac{\ds \eta_k \gamma_k}
{\ds \!\!\sum_{j=1}^K {\eta_j \xi_{kj}} + \eta_R \zeta_{k,R}\left(\phi,\theta\right)\! + \!\sigma^2_z } \geq t \; \forall k \\
& \ds \sum_{k=1}^K \eta_{k} + \eta_R \leq \ds \frac{P_{\rm max}}{MN}\\
& \ds \frac{\eta_R \norm{\mathbf{a}\left(\phi,\theta\right)\mathbf{a}^H\left(\phi,\theta\right)\mathbf{w}_R\left(\phi,\theta\right)}^2}{\ds \sum_{k=1}^K \eta_k \norm{\mathbf{a}\left(\phi,\theta\right)\mathbf{a}^H\left(\phi,\theta\right)\mathbf{w}_k}^2 } \geq \rho^*  
\end{array}\right.
\label{Feas_program}
\end{equation}

\IF {Problem \eqref{Feas_program} is feasible}

\STATE $t_{\rm min}=t$

\ELSE 

\STATE $t_{\rm max}=t$

\ENDIF

\ENDWHILE

\end{algorithmic}

\end{algorithm}

\begin{table}[]
\centering
\caption{Simulation Parameters}
\label{Sim_Par}
\begin{tabular}{|p{1.5cm}|p{1.5cm}|p{4cm}|}
\hline
\textbf{Name}                   & \textbf{Value} & \textbf{Description}                                                                                                                \\ \hline
$f_c$                           & 3 GHz          & carrier frequency                                                                                                                   \\ \hline
$M$                             & 512    & number of subcarriers
 \\ \hline
$N$                             & 14    & number of OFDM symbols
 \\ \hline
$\Delta_f$                            & 30 kHz    & subcarrier spacing
 \\ \hline
 $B=\Delta_f M$                            & 15.36 MHz    & system bandwidth
 \\ \hline
  $T_0$                            & 0.357 $\mu$s    & OFDM symbol duration
 \\ \hline
$K$                             & 10          & number of users in the cellular system 
\\ \hline
$F$                             & 9 dB           & noise figure at the receiver                                                                                                        \\ \hline
$\mathcal{N}_0$                 & -174 dBm/Hz    & power spectral density of the noise                                                                                                 \\ \hline
\end{tabular}
\end{table}

\begin{figure*}[!t]
\centering
\includegraphics[scale=0.5]{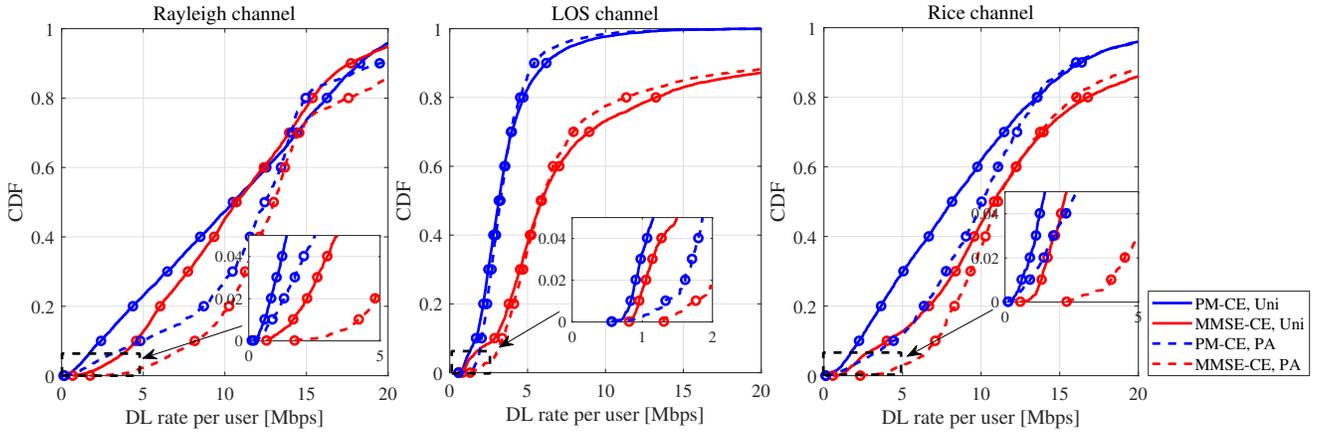}
\caption{CDFs of DL rate per user using the PBR approach, with uniform (Uni) and proposed power allocation strategy (PA). Rayleigh channel, LoS channel and Rice channel, RCR $=3$ dB and $N_{A,y}\times N_{A,z}=10 \times 10$.}
\label{Fig:Rate_channels_PBR_ZFR}
\end{figure*}

\section{Numerical results} \label{Numerica_results}

The parameters for the simulation setup are reported in Table \ref{Sim_Par}. We assume that the users of the communication system are randomly located on the $(x,y)$ plane with $x$ in $[10,100]$ m and $y$ in $[-50,-10] \cup [10, 50]$, with heighs 1.65 m. The height of the radar-BS is 15 m. 
For the Rayleigh channel model in Eq. \eqref{Rayleigh_channel}, we follow the three slope path loss model in \cite{buzzi_CFUC2017} and we assume uncorrelated shadow fading. For the LoS channel in Eq. \eqref{LOS_channel}, the path-loss follows the model in \cite[Table B.1.2]{3GPP_36814_GUE_model}, while  for the Rice channel in Eq. \eqref{Rice_channel} we use again the model in \cite{buzzi_CFUC2017} and  the LoS probability is evaluated following \cite{3GPP_36873}. The quantity $\alpha_T$ in Eq. \eqref{radar_channel} containing the target reflection coefficient and the path-loss is modeled as
$\alpha_T=G\sqrt{\frac{\zeta}{L_{\tau}}}$, 
where $G=10 \log_{10}(N_A)$ dB is the radar-BS antenna gain, $\zeta=0.1253 \text{m}^2$ is the target radar cross section (RCS)\footnote{The RCS of a common unmanned aherial vehicle (UAV) \cite{Chenchen_2016_Drones} has been chosen.} and 
$L_{\tau}=\frac{(4 \pi)^3}{\lambda^2}\left( \frac{c \tau}{2}\right)^4$.
We define the Radar-Communication-Ratio (RCR) as  
$
\text{RCR}=P_R/P_{\rm DL}.
$
The scanning area of the radar system  extends for $[-60, 60]^o$ in azimuth and for $[10, 80]^o$ in elevation. In the following results we compare the performance obtained with the proposed power allocation (PA) in Section \ref{Power_all_section} with the uniform power allocation (Uni). In the case of Uni we assume $\eta_k=P_{\rm DL}/(K M N)$, with $P_{\rm DL}= 2 $ W the radar-BS power budget used for communication tasks. The SIR contraint in Eq. \eqref{Prob:cMinRate}, $\rho^*$, is the RCR.

\begin{figure}[!t]
\centering
\includegraphics[scale=0.5]{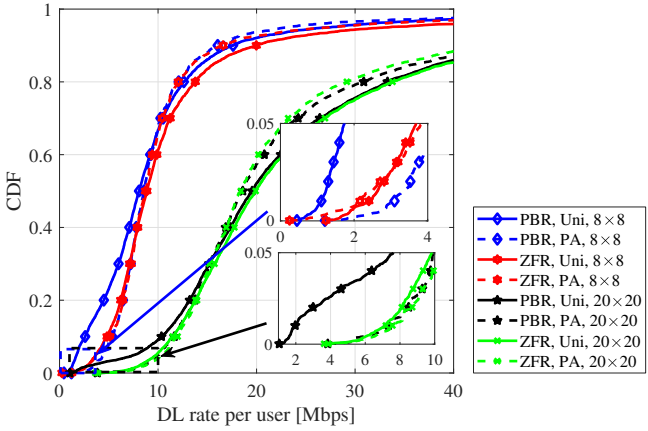}
\caption{CDF of DL rate per user using the PBR and ZFR approaches, with uniform (Uni) and proposed power allocation strategy (PA). Rayleigh channels, RCR $=3$ dB, two values of $N_{A,y}\times N_{A,z}$.}
\label{Fig:Rate_AntennaConf_PBR_ZFR}
\end{figure}

\begin{figure}[!t]
\centering
\includegraphics[scale=0.42]{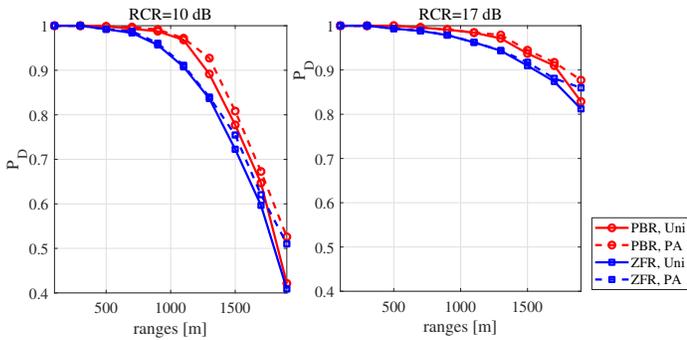}
\caption{Probability of detection versus range using the PBR and ZFR approaches, with uniform (Uni) and proposed power allocation strategy (PA). Rayleigh channel for the users, two values of RCR, $N_{A,y}\times N_{A,z}=10 \times 10$.  }
\label{Fig:Detection_probability}
\end{figure}

Fig. \ref{Fig:Rate_channels_PBR_ZFR} reports the cumulative distribution functions (CDFs) of the downlink (DL) rate per user obtained in the Uni and PA cases for the three channel models  discussed in Section \ref{Channel_model_section} with RCR=3 dB, with PM-CE and MMSE-CE, and assuming the PBR approach in Eq. \eqref{pure_radar_direction} for the radar task. 
Results show that performance with PA is better with respect to the one obtained in case of Uni in terms of fairness, as shown in the zoomed part of each subfigure. In particular, it is clearly seen that the PA algorithm produces a clear improvement of the lower tail of the CDF of the rates. The figure also permits assessing the impact of the CE techniques on the system performance, in particular the MMSE-CE offers better performance with respect to the PM-CE, because the former exploit the knowledge of the second order statistic of the users' channels. 
Fig. \ref{Fig:Rate_AntennaConf_PBR_ZFR} reports the DL rate per user in the cases of Rayleigh channels for the users, fixed RCR=3 dB, and for two antenna configurations at the radar-BS in the case of MMSE-CE at the radar-BS using both the PBR and the ZFR in Eq. \eqref{ZF_radar}. The figure permits assessing the beneficial impact obtained increasing the antenna array size. 
Clearly, a larger antenna size permits on one hand capturing more energy when receiving, and, on the other, producing narrower beams when transmitting, which eventually results in lesser interference to the mobile stations.
Additionally, we can note that the gain in performance is better in the case of PBR with respect to the one in the case of ZFR.
In Fig. \ref{Fig:Detection_probability}, we report the probability of detection $P_D$ versus the target distance, using Rayleigh channel for the users and two values of RCR assuming a false alarm probability of $10^{-2}$. It can be seen, as expected, that the detection performance in the case of ZFR is worse than that achieved with PBR: indeed, nulling the interference between the radar signal and the users has a negative impact on the shape of the beam used for target detection. Additionally, we can see that the performance obtained with the PA outperforms the case with Uni, so using the proposed power allocation strategy brings also some benefits in terms of detection capabilities of the system, possibly due to the radar SIR constraint present in the formulated optimization problem. 

\section{Conclusions} \label{Conclusions}
The paper has analyzed the case in which a radar-BS equipped with massive MIMO arrays is used for joint communications and sensing tasks. Building upon the system model and the related signal processing algorithms introduced in reference \cite{Buzzi_Asilomar2019}, a power allocation strategy that maximizes the fairness across the users on the ground with a constraint on the SIR on the radar task has been proposed and numerically assessed. Further research on this topic may be focused on the problem of devising advanced signal processing algorithms for increased performance and the capability of tracking trajectories of flying targets. This forms the object of current investigation.

\ifCLASSOPTIONcaptionsoff
  \newpage
\fi
\bibliographystyle{IEEEtran}

\bibliography{MyReference}

\end{document}